\documentstyle[preprint,aps,epsf]{revtex}
\tightenlines
\def\Journal#1#2#3#4{{#1} {\bf #2}, #3 (#4)}
\def\ADP{{\em Adv. in Phys.}}
\def\AP{{\em Ann. Phys.}}

\def\NPB{{\em Nucl. Phys.} B}

\def\PRL{\em Phys. Rev. Lett.}

\def\PRA{{\em Phys. Rev.} A}

\def\PRD{{\em Phys. Rev.} D}

\def\JMP{{\em J. Math. Physics.}}
\def\RMP{{\em Rev. Mod. Phys.}}
\def\PREPC{{\em Phys. Rep.} C}

\newcommand{\be}{\begin{equation}}
\newcommand{\ee}{\end{equation}}
\newcommand{\bea}{\begin{eqnarray}}
\newcommand{\eea}{\end{eqnarray}}

\newcommand{\hf} {{1\over2}}
\newcommand{\nonu}{\nonumber\\}
\def\dk{\Delta k}

\begin{document}
\title{Instability Induced Renormalization}

\author{Jean Alexandre\thanks{alexandr@lpt1.u-strasbg.fr}}
\address{Laboratory of Theoretical Physics, Louis Pasteur University\\
3 rue de l'Universit\'e 67087 Strasbourg, Cedex, France}

\author{Vincenzo Branchina\thanks{branchin@lpt1.u-strasbg.fr}}
\address{Laboratory of Theoretical Physics, Louis Pasteur University\\
3 rue de l'Universit\'e 67087 Strasbourg, Cedex, France}

\author{Janos Polonyi\thanks{polonyi@fresnel.u-strasbg.fr}}
\address{Laboratory of Theoretical Physics, Louis Pasteur University\\
3 rue de l'Universit\'e 67087 Strasbourg, Cedex, France\\
and\\
Department of Atomic Physics, L. E\"otv\"os University\\
Puskin u. 5-7 1088 Budapest, Hungary}
\date{\today}
\maketitle
\begin{abstract}
It is pointed out that models with condensates have nontrivial
renormalization group flow on the tree level. The infinitesimal
form of the tree level renormalization group equation is obtained
and solved numerically for the $\phi^4$ model in the symmetry
broken phase. We find an attractive infrared fixed point that
eliminates the metastable region and reproduces the Maxwell
construction.
\end{abstract}

We have two systematic nonperturbative methods to handle  multi-particle 
or quantum systems, the saddle point approximation and the 
renormalization group. Our goal in this letter is to combine these two
apparently independent approximation methods in order to
obtain a better understanding of the instabilities and
first order phase transitions.

We start with the path integral,
\be
Z=\int \left[{\cal D}\phi\right]e^{-{1\over\hbar}S[\phi]},\label{partfc}
\ee
over the configurations $\phi(x)$. We assume the presence of
UV and IR cutoffs thus the dimension of the domain of 
integration is large but finite.

The saddle point approximation to (\ref{partfc}) coincides with
its perturbative expansion when the saddle points are trivial, $\psi=0$.
In this case the elementary excitations,
the quasiparticles, are characterized by the eigenfunctions
$\phi_n(x)$ of the inverse propagator, 
$G^{-1}(x,y)=\delta^2S[\phi]/\delta\phi(x)\delta\phi(y)
_{\Big\vert\phi=\psi}$.
The perturbation expansion is applicable for 
fixed values of the cutoffs if the restoring force of the fluctuations to 
$\phi(x)=\psi(x)=0$ is nonvanishing, i.e. the eigenvalues of the inverse 
propagator, $\lambda_n$, are positive. 
When the absolute minimum of the action
is reached at $\phi(x)=\psi(x)\not=0$ then we follow the strategy of 
the saddle point expansion and the elementary excitations are
the fluctuations around $\psi(x)\not=0$. The saddle point 
expansion is applicable as long as $\lambda_n>0$.
The zero modes, the elementary excitations
with $\lambda_n=0$, correspond to continuous symmetries and
are integrated over exactly. 
If the inverse propagator has too many small eigenvalues
the saddle point expansion breaks down and strong fluctuations
develop around $\psi$ as we remove one of 
the cutoffs. This possibility brings us to the 
renormalization group method which is supposed to deal with such
problems \cite{wilsrg}.

The basic idea of the renormalization group
is the subsequent integration in (\ref{partfc}),
\be
Z=\int d\phi_1\int d\phi_2\cdots
\int d\phi_N e^{-{1\over\hbar}S_N[\phi]},\label{partfcrg}
\ee
where the field is expanded according to a suitable chosen
basis, $\phi(x)=\sum_{n=1}^N\phi_n\Phi_n(x)$,
and $N<\infty$ plays the role of the UV cutoff.
The effective theory for the first N' modes
is defined through the effective action $S_{N'}$ by the help
of the blocking transformation for $\Delta N=N-N'$ number of modes,
\be
e^{-{1\over\hbar}S_{N'}[\phi]}=\int d\phi_{N'+1}
\cdots\int d\phi_Ne^{-{1\over\hbar}S_N[\phi]}.\label{blocking}
\ee
The elimination step is usually followed 
by a rescaling in order to restore the cutoff to its
original value. This rescaling, together with its result, the anomalous
dimension, is an important device to distinguish the trivial (tree-level) and 
the dynamical (loop-level) scale dependence.

The fluctuations of the IR modes with $n<N$ can be
described within the effective theory given by the action
\be
S_N[\phi]=\sum_jg_j(N)s_j[\phi,\partial_\mu\phi,\Box\phi,...],\label{effact}
\ee
where $s_j$ is a suitable chosen complete set of local functionals
without keeping the UV modes $n>N$ present explicitly. Thus
the IR modes decouple from the UV ones
since the correlations generated by the latters are contained 
in the numerical value of the effective coupling constant $g_j(N)$.
This trivial observation gives rise a powerful
approximation scheme by the truncation of (\ref{effact}) when
{\em any} IR field configuration can be used to reconstruct
the {\em same} overdetermined effective action.
In the infinitesimal form of the renormalization group method, \cite{wh}, 
where $\Delta N<<N$ the small parameter of the loop expansion in 
(\ref{blocking}) is $\hbar\Delta N/N$, thus
we expect that the saddle point approximation
is applicable in (\ref{blocking}). The Wegner-Houghton equation \cite{wh}
is obtained by eliminating the plane waves within 
the shell $k-\dk<|p|<k$ of the momentum space,
\bea
{\partial S_k[\phi]\over \partial k}
&=&{\hbar\over\dk}\left[\hf{\mathrm Tr}_{k,\dk}{\delta^2S[\phi]\over\delta\phi\delta\phi}
-{1\over\hbar}{\delta S[\phi]\over\delta\phi}
\cdot\left({\delta^2S[\phi]\over\delta\phi\delta\phi}\right)^{-1}
\cdot{\delta S[\phi]\over\delta\phi}\right]\nonu
&&-{\eta+d\over2}\partial_\mu\phi\cdot{\delta S[\phi]\over\delta\partial_\mu\phi}
+{2-\eta-d\over2}\phi\cdot{\delta S[\phi]\over\delta\phi},\label{whe}
\eea
where the trace and the "$\cdot$" operation is taken in the subspace of the
eliminable modes. The second line generates the rescaling assuming the 
anomalous dimension $\eta$ (input) in dimension $d$. 
The second order approximation is used for the
action as the functional of the eliminable field variables which is applicable
only if the saddle point amplitude is $O(\hbar)$. Our interest in this work
is the case where there is a saddle point during the elimination of the modes
and we shall see that there is no gurantee in general that the saddle point 
remains $O(\hbar)$ during the evolution. Thus
we need a more reliable evolution equation which is applicable without
imposing the condition $O(\hbar)$ on the saddle point amplitude.

Systems with instabilities display fluctuations with large 
amplitudes which can be taken into account by means of the
saddle point approximation. In order to demonstrate the importance 
of the saddle point during the blocking 
on the renormalized trajectory we shall consider the $\phi^4$ model 
in the symmetry broken phase in a box with linear size $L$ 
and with periodic boundary conditions. The bare action of this model is
$S_\Lambda=\int d^dx[{1\over2}(\partial_\mu\phi)^2+U_\Lambda(\phi)]$
where $U_\Lambda(\phi)=-m^2/2\phi^2+g\phi^4/4!$ ($\Lambda$ is the UV cutoff)
and we search for the vacuum in the presence of the constraint
$L^{-d}\int d^dx<\phi(x)>=\Phi$. The constrained vacuum displays
spinodal instability or metastability when it is unstable
against fluctuations with infinitesimal or finite amplitude, respectively.
The spinodal instability can be detected by local methods, mode by mode
inspection. In fact, each fluctuation of the form
\be
\psi_k(x)=\tilde\psi_k e^{ik_\mu x_\mu}+\tilde\psi_k^\star e^{-ik_\mu x_\mu}
=2\rho_k\cos(k_\mu x_\mu+\alpha_k)\label{oplawe}
\ee
with infinitesimal amplitude, $\rho_k\approx0$, represents an
independent unstable mode of the tree level theory when
$p^2<m^2-g\Phi^2/2$. The 
Legendre transform of the tree level theory suggests
that the vacuua with $2m^2/g<\Phi^2<6m^2/g$ are metastable.
The vacuum is stable when $\Phi^2>6m^2/g$. 
It is difficult to identify the possible instabilities
of the true vacuum where the local, mode by mode analysis
is unreliable. This is because the decay of the unstable
vacuum udergoes large amplitude modifications either 
in the metastable or the spinodal instable region. 

We show now that the renormalization group can be used to deal with 
the modes of the unstable vacuum in a simple one by one manner.
The blocking transformation we employ consists of the
elimination of the modes with momentum $k-\dk<|p|<k$, where 
$k$ is the current cutoff and $\dk$ is a momentum parameter 
smaller than any characteristic momentum scale of the system.
The effective coupling constants
can be defined in the leading order of the gradient expansion 
by $U_k(\phi)=\sum_jg_j(k)/j!\phi^j$. 
Since any infrared field configuration should yield the same
effective action we take the simplest choice, a homogeneous
field $\phi(x)=\Phi$. 

The loop expansion for (\ref{blocking}) yields the general result
$S_{N'}=\sum_j\hbar^jS^{(j)}_{N'}$. The perturbation expansion
retains the term $O(\hbar^j)$ with $j>0$ and the tree
level, $O(\hbar^0)$ contribution represents a non-perturbative
piece what must be considered {\em before} the perturbative
pieces are taken into account. Since the tree level expressions are 
different from the loop corrections the tree level renormalization,
if exists, consists of essentially new and different expressions than the
loop contributions which have been considered so far \cite{all}.
Our constraint generates non-trivial saddle points and we consider here the
leading order, tree level renormalization only. In this
approximation the naive scaling is correct ($\eta=0$, $Z=1$), and we ignore the 
rescaling step of the renormalization group method for the sake of simplicity.

The saddle point of the blocking transformation can be written in the form 
$\psi_k(x)=\sum'\tilde\psi_p e^{ipx}$ where the prime
denotes the summation for $k-\dk<|p|<k$. The blocked potential is given by
\be\label{blpot}
L^dU_{k-\dk}[\Phi]=-\hbar\ln\left[\int[{\cal D}\eta]e^{-\frac{1}{\hbar}S_k[\Phi+\eta]}\right]
=\min_{\{\psi\}}\left[
\int d^dx\left(\frac{1}{2}(\partial_\mu\psi)^2+
U_k(\Phi+\psi)\right)\right]+O(\hbar)
\ee
where the fluctuation $\eta$ contains only the modes within the shell $[k-\dk,k]$.
It is worthwhile noting that (\ref{blpot}) reduces to the usual 
local potential approximation of the Wegner-Houghton equation 
\cite{locpot} if the saddle point vanishes.
There might be several minima $\psi$ in which case one should sum over them. 
We retain only the {\it single plane wave} saddle points, (\ref{oplawe}).
The motivation of this rather drastic approximation is the assumption that 
the saddle point (\ref{oplawe}) seems to be optimal from the point of view of the
energy-entropy ballance. To see this first note that when $\dk\to0$ the saddle point
is non-vanishing in the momentum space on the sphere with radius $k$
only. As far as the saddle point on this sphere is concerned, the introduction
of other additional plane waves would increase the kinetic energy. 
The phase $\alpha_k=\alpha_{-k}$ is a zero mode for each plane waves,
corresponding to the translation invariance. 
The more involved saddle points built by several plane waves have the same
single translational zero mode. Thus the kinetic energy-entropy balance
is optimized for the plane waves. The resulting tree-level blocking relation is
\bea
U_{k-\dk}(\Phi)&=&\min_{\{\rho\}}\left[
k^2\rho^2+\frac{1}{2}\int_{-1}^1 du~
U_k(\Phi+2\rho\cos(\pi u))\right]\nonumber\\
&=&k^2\rho^2_k+\frac{1}{2}\int_{-1}^1 du~U_k(\Phi+2\rho_k\cos(\pi u)).
\label{sblpot}
\eea

We followed the tree level evolution of the potential
by performing numerically the minimization in (\ref{sblpot}) at each
blocking step. The most interesting aspect of the result is
that, starting from $k=\sqrt{m^2}-\dk$,
the saddle points $\psi_k$ are nonvanishing 
(and consequently the potential receives a 
nontrivial renormalization) for
$0<\Phi^2<\Phi^2_{vac}(k)=6(m^2-k^2)/g$. Thus the mode
coupling provided by our successive elimination method 
extends the spinodal instability over the vacuua which are
seen as metastable by the tree level Legendre 
transformation. This can happen because once the tree level
contributions are found at modes with $p^2<m^2$ 
for certain values of $\Phi^2$ the
saddle point contributions renormalize the potential
in such a manner that spinodal instability occures for other,
larger values of $\Phi^2$.

Other important results of the numerical analysis are in order:\\
(i) the saddle point amplitude is a linear function
of the field, $2\rho_k=-\Phi+\Phi_{vac}(k)$;\\
(ii) we find the potential
\be\label{rpot}
U_k(\Phi)=-\frac{1}{2}k^2\Phi^2-\frac{3}{2g}(m^2-k^2)^2
\ee
whenever the saddle points are nontrivial, i.e. in
the unstable region, see Fig.1, where the evolution of the potential
with $m^2=0.1$ and $g=0.2$ is shown;\\
(iii) the blocking converges as $\dk\to0$ 
despite the apparent absence of $\dk$ in (\ref{sblpot});\\
(iv) we checked that the previous results are {\it universal}
with respect to the choice of the symetry breaking bare
potential.

It is possible to understand the result (iii) if we write
(\ref{sblpot}) for two subsequent steps and take the difference:
\bea\label{fdre}
U_{k-2\dk}(\Phi)&=&U_{k-\dk}(\Phi)-\dk\left[
2k\rho_k^2+2k^2\rho_k\partial_k\rho_k
+\frac{1}{2}\int_{-1}^1du~\partial_k 
U_k\left(\Phi+2\rho_k\cos(\pi u)\right)\right.\nonumber\\
&+&\left.\int_{-1}^1du~\cos(\pi u)~\partial_k\rho_k~
\partial_\phi U_k\left(\Phi+2\rho_k\cos(\pi u)\right)\right]
+{\cal O}(\dk)^2,
\eea
which shows explicitely that the correction to the potential
due to one blocking step is proportional to $\dk$.  
It is also worth remarking that the potential (\ref{rpot})
is a solution of the equation (\ref{fdre}), as can easely be 
verified.

Few remarks are in order at this point.
(a) The disappearance of the metastable region is in agreement
with the finding of ref. \cite{bist}. The saddle points
of the static problem reflect the spontaneous droplet 
formation dynamics \cite{instab}: Suppose that the
stable vacuum bubble in a sphere of radius $R$ and
created in the false vacuum has the energy
$E(R)=4\pi R^2\sigma-4\pi R^3\Delta E/3$ where $\sigma$ and
$\Delta E$ stand for the surface tension and the (free)energy
difference between the two vacua. The critical
droplet size, $R_c=\sigma/\Delta E$, beyond which
the droplets grow can be identified by the inverse of 
the highest momentum of the condensate, 
$R_c^2=6/g(\Phi_{vac}^2(0)-\Phi^2)$.
This relation establishes a connection between the static 
and the dynamical properties. 
(b) The last term in the right hand side of (\ref{rpot}) 
was added by hand to achieve a continuous matching of the 
partition function at the instability. As a result, (\ref{rpot}) 
reproduces the Maxwell construction for the effective potential 
$V_{eff}(\Phi)=U_{k=0}(\Phi)$ \cite{maxw}, i.e.
it remains unchanged (at the tree level) in the stable
region, $\Phi^2>\Phi^2_{vac}(0)$, and turns out to be constant and continuous
for $\Phi^2\le\Phi^2_{vac}(0)$.  
(c) The potential (\ref{rpot}) contains only one coupling constant,
$m^2(k)=\partial^2_\phi U_k(0)=-k^2$. The corresponding dimensionless 
coupling constant, $\tilde m^2(k)=m^2(k)/k^2=-1$, is trivially
renormalization group invariant. In other words
we recover the usual scaling laws in the stable
region, while in the unstable region 
the potential is a fixed point given by (\ref{rpot}).
(d) Similar tree level potential has been found in the N-component $\phi^4$ model
with smooth cutoff \cite{riwe} where the would be unstable and the stable regime
join and the naive metastable region is wiped out by the Goldstone modes. 
While the possibility of having a homogeneous $\psi_k^2(x)$ is an
essential part of the argument, our result shows the more general origin
of the potential. Furthermore note that the keeping of a single plane wave mode 
as a saddle point is not justified when a smooth cutoff is used. The analytical
structure of the loop corrections in the vicinity of the instability 
indicates the same potential as well, \cite{tetr}. The availability of
such a diverse derivations and the microscopic dynamics-independent
form suggests a more fundamental origin of this potential. We believe that
underlying reason of (\ref{rpot}) is the Maxwell construction and the
actual form can easiest be identified by means of the renormalization group
method where the dynamics of each mode can be dealt with individually.
In fact, one can show that the differentiability
of the renormalization group flow, i.e. the existence of the beta functions,
and the nonvanishing of the saddle points as $\dk\to0$
requires the form (\ref{rpot}). To see this consider the finite difference
\be
{U_{k-\dk}(\Phi)-U_k(\Phi)\over\dk}={k^2+\partial_\Phi^2U_k(\Phi)
\over2L^d\dk}\int dx\psi_k^2(x)+{\partial_\Phi^3U_k(\Phi)
\over6L^d\dk}\int dx\psi_k^3(x)+\cdots+O(\hbar),
\ee
where we expanded (\ref{blpot}) in the saddle point, $\psi_k(x)$. The
convergence of the left hand side requires either the smallness of the
saddle point, $\psi_k=O(\dk^{1/2})$, or $k^2+\partial_\Phi^2U_k(\Phi)=
O(\dk)$ and $\partial^n_\Phi U_k(\Phi)=O(\dk)$. Since the
saddle point is nonvanishing the form (\ref{rpot}) follows.
Note the essential differences between the tree and the loop level 
renormalization: The former is nonalytic in $g$ and lacks the usual 
logrithms of the latter.
(e) The result (i) can be obtained by assuming (\ref{rpot})
in the unstable region and requiring continuity 
of $\partial_\Phi U_k(\Phi)$ in $\Phi$. 

The saddle point approximation includes the softest fluctuations,
the zero modes, and one can obtain nontrivial contributions
to the correlation functions at $O(\hbar^0)$. In fact, let us insert the
product of field variables in the integrand of (\ref{partfcrg})
and follow the
successive elimination of the modes in the path integral. The resulting
tree level expression for the $2n$-point function on the background field
$\Phi$ is 
\bea
G^{(0)}_\Phi(p_1,\cdots,p_{2n})&=&
{\int{\cal D}[\hat k]\int{\cal D}[\alpha]\prod_{j=1}^{2n}
\tilde\psi_{k_j}(p_j)\over
\int{\cal D}[\hat k]\int{\cal D}[\alpha]}\nonu
&=&\sum_P\prod_{j=1}^n\delta(p_{P(2j)}+p_{P(2j-1)})
{d(2\pi)^d\over\Omega_dk^d(\Phi)}\rho^2_{p_j}\label{prop}
\eea
where $\tilde\psi_k(p)=\int d^d e^{-ipx}\psi_k(x)$, $\Omega_d$
stands for the solid angle,
we integrate over the zero modes characterized by the
unit vector $\hat k(k)$ corresponding for each value of the
cutoff, $k$, and the phase $\alpha(p)$.
The sum in the second equation is over the permutations $P$ of the 
field variables and $k^2(\Phi)=m^2-g\Phi^2/6$.
Whenevere we have an odd number of fields or $\Phi^2\ge\Phi^2_{vac}(0)$
or one of the $p_j$ is such that $p_j^2>m^2-g\Phi^2/6$, $G^{(0)}=0$.
The integration over the zero modes
is reminiscent of the integration over the possible rearrangements
of the domain walls in the mixed phase and restores the translation invariance
of the correlations. The two point function $G^{(0)}_\Phi(p_1,p_2)
=2{d(2\pi)^d\over\Omega_dk^d(\Phi)}\rho^2_{p_1}\delta(p_1+p_2)$ 
shows that the Fourier transform of 
$\tilde\psi_p$ can be interpreted as the domain wall structure for a given
choice of the zero modes. 

The method put forward in this letter, the use of the saddle
point approximation for the blocking transformation, can be used
to handle some of the systems where large amplitude, inhomogeneous
fluctuations are present. We studied here an Euclidean system describing
the equilibrium situation. Whenever the time dependence far from 
equilibrium has nontrivial semiclassical limit this method can be
extended to include the real time dependence, \cite{rqm}.
Another natural continuation of this work is the inclusion of the
loop corrections in addition to the tree level pieces.
The correlation functions computed in such a manner
include the fluctuations in a systematic manner which is an
improvement compared to ref. \cite{laba} where the master equation
was used to describe the most probable values of the correlation functions.

\begin{figure}
\begin{minipage}{5cm}
	\epsfxsize=5cm
	$$
	\epsfysize=5cm
	$$
	\centerline{\epsfbox{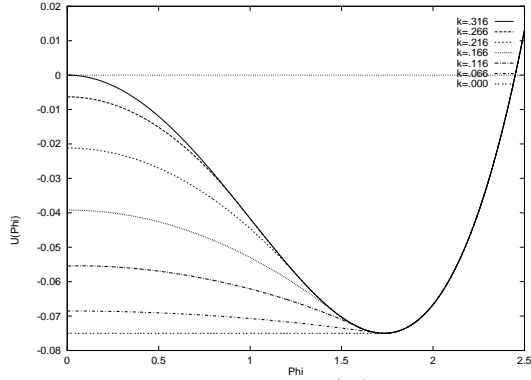}}
\end{minipage}
\caption{The potential $U_k(\Phi)$ for different values of
$k$ showing the RG evolution towards the Maxwell effective potential.}
\end{figure}

\acknowledgments
We thank Daniel Boyanowski for a useful discussion.


\begin{references}
\bibitem{wilsrg}K. Wilson and J. Kogut, \Journal{\PREPC}{12}{75}{1974};
K. Wilson, \Journal{\RMP}{47}{773}{1975}. 
\bibitem{wh} F. J. Wegner, A. Houghton, \Journal{\PRA}{8}{40}{1973}. 
\bibitem{grg} J. Alexandre, V. Branchina, J. Polonyi, 
\Journal{\PRD}{58}{016002}{1998}.
\bibitem{all} 
J. Polchinski, \Journal{NPB}{231}{269}{1984};
A. Hasenfratz, P. Hasenfratz, \Journal{\NPB}{295}{1}{1988};
C. Wetterich, \Journal{\NPB}{352}{529}{1991}.
\bibitem{locpot} J. F. Nicoll, Chang, Stanley, \Journal{\PRL}{33}{540}{1974};
A. Hasenfratz, P. Hasenfratz, \Journal{\NPB}{270}{685}{1986};
C. Wetterich, \Journal{\NPB}{352}{529}{1991};
S. B. Liao, J. Polonyi, \Journal{\AP}{222}{122}{1993};
T. R. Morris, \Journal{\NPB}{458}{477}{1996}.
\bibitem{instab} J. S. Langer, in {\em Solids Far From Equilibrium},
ed. C. Godreche, Cambridge Univ. Press, 1992.
\bibitem{bist} K. Binder, D. Staufer, \Journal{\ADP}{25}{343}{1976}.
\bibitem{maxw} Y. Fujimoto, L. O'Raifeartaigh, G. Parravicini, 
\Journal{\NPB}{212}{268}{1983}; R. W. Haymaker, J. Perez-Mercader,
\Journal{\PRD}{27}{1948}{1983}; C. M. Bender, F. Cooper, \Journal{\NPB}{224}
{403}{1983}; F. Cooper, B. Freedman, \Journal{\NPB}{239}
{459}{1984}; 
M. Hindmarch, D.Johnson, \Journal{\JMP}{A19}{141}{1986};
V. Branchina, P. Castorina, D. Zappal\`a, \Journal{\PRD}{41}{1948}{1990};
K. Cahill, \Journal{\PRD}{52}{4704}{1995}.
\bibitem{riwe} A. Ringwald, C. Wetterich, \Journal{\NPB}{334}{506}{1990}.
\bibitem{tetr} N. Tetradis, C. Wetterich, \Journal{\NPB}{383}{197}{1992};
N. Tetradis, \Journal{\NPB}{488}{92}{1997}.
\bibitem{rqm} J. Polonyi, \Journal{\AP}{55}{333}{1996}.
\bibitem{laba} J. S. Langer, M. Bar-on, H. Miller, \Journal{\PRA}{11}{1417}{1975}.
\end{references}
\end{document}